# Morphology-driven electrical and optical properties in graded hierarchical transparent conducting Al:ZnO

P. Gondoni[1], P. Mazzolini[1,2], A. M. Pillado Pérez[1], V. Russo[1], A. Li Bassi[1,2], C. S. Casari[1,2]

[1]Dipartimento di Energia and NEMAS – Center for NanoEngineered Materials and Surfaces, Politecnico di Milano
Via Ponzio 34/3, 20133 Milano, Italy

[2]Center for Nano Science and Technology @Polimi, Istituto Italiano di Tecnologia
Via Pascoli 70/3, 20133 Milano, Italy

**Abstract**

Graded Al-doped ZnO layers, constituted by a mesoporous forest-like system evolving into a compact transparent conductor, were synthesized by Pulsed Laser Deposition with different morphology to study the correlation with functional properties. Morphology was monitored by measuring the resulting surface roughness and its effects on electrical conductivity – especially carrier mobility, which significantly decreases with increasing roughness – allow to discuss the limitations in conduction mechanisms. Significant changes in light scattering capability due to variations in morphology are also investigated and discussed to study the correlation between morphology and functional properties.

1. **Introduction**

    The well-established importance of transparent conducting materials (TCMs) in optoelectronic and energy conversion devices has led to the development of highly performing materials, the most widely employed being Indium-Tin Oxide (ITO). [1–4] Due to the scarcity and high price of indium, ITO needs to be replaced by cheaper and more abundant materials: Aluminum doped Zinc Oxide (AZO) is among the most promising candidates and has been investigated for several years. [5–7]

    However, as the field of photovoltaics progresses towards hybrid and organic-based flexible devices, the optical transparency and electrical conductivity of TCMs need to be accompanied by more advanced properties, such as large effective surface area, light trapping and management, mechanical flexibility and low temperature synthesis. [8–10]

By employing Pulsed Laser Deposition (PLD) as a deposition technique, it is possible to obtain control of local structure, multi-scale morphology and stoichiometry for a number of oxides even at room temperature [11–14] and thus to exploit this versatility to obtain with AZO the functional properties indicated above [15–17].

We have recently developed a functionally graded AZO structure, which is able to combine optical transparency and electrical conductivity with effective light management. [18] This structure is constituted by a bottom mesoporous layer designed to maximize light scattering, smoothly evolving into a compact transparent conducting AZO layer on top. The realization of this architecture is based on the effects of background pressure on the PLD process: at pressures of the order of 100 – 200 Pa $O_2$ deposition takes place in the form of a porous hierarchical assembly of nanosized clusters, whose formation is favored by collisions with the background gas; as the background pressure is decreased (0.1 – 10 Pa $O_2$), compact films are obtained, as discussed in previous works.[17], [19], [20]

In the development of this functionally graded material [18] we identified a significant interconnection between the porosity and thickness of the bottom porous layer, the surface morphology of the top compact layer surface and the resulting functional properties (i.e. electrical conductivity and light scattering).

The aim of this work is to study in detail the correlation between the morphology of graded AZO (in particular, its surface morphology, quantified by surface roughness values) and its electrical and optical properties. We will show how it is possible to investigate the morphology-related mechanisms behind electrical conductivity (in particular, carrier mobility), visible optical transparency and haze (i.e. scattered to total transmitted intensity ratio).

2. **Material and methods**

AZO films were grown by PLD at room temperature on Si(100) and soda-lime glass substrates. A solid 2%wt. $Al_2O_3$:ZnO target was ablated by a ns pulsed laser (Nd:YAG 4[th] harmonic, λ=266 nm,

$f_p$=10 Hz) in the presence of oxygen as a background gas. At a 50 mm target-to-substrate distance the laser energy density on the target was ≈ 1 Jcm$^{-2}$. The O$_2$ pressure was varied during deposition to control morphology: porous layers were grown at pressures in the range from 100 to 110 Pa, then the pressure was uniformly decreased (≈ 6.5 Pa/min) to obtain a buffer layer evolving into a compact layer grown at 2 Pa O$_2$. The background pressure for the compact layer was chosen to obtain optimized electrical and optical properties [19]. Deposition times and pressures of the bottom layers were adjusted for different samples in order to control surface roughness of the graded film. Scanning Electron Microscopy (SEM) images were acquired with a Zeiss SUPRA40 Field-Emission SEM. Atomic Force Microscopy (AFM) measurements were performed in tapping/noncontact mode with a Thermomicroscopes Autoprobe CP II using Veeco RTESPA tips (spring constant 20-80 N/m, resonance frequency 200-300 kHz). Grain size analysis was carried out with the SPIP$^{TM}$ software. Resistivity and Hall coefficients were measured in the Van der Pauw configuration with a Keithley K2400 as a current source (from 100 nA to 10 mA), an Agilent 34970A voltage meter and a 0.57 T Ecopia permanent magnet.

Optical transmittance and haze spectra were acquired with a UV–vis–NIR PerkinElmer Lambda 1050 spectrophotometer with a 150 mm diameter integrating sphere. The spectra were normalized to correct for substrate (glass) contribution by setting to 1 the intensity at the glass/film interface.

## 3. Results

### 3.1. Control of surface roughness

Functionally graded AZO layers were grown in different conditions in order to obtain different morphologies, characterized by increasing surface roughness values. The deposition parameters which were varied were the oxygen pressure and time employed for the growth of the porous scattering layer, leaving the buffer and compact layers unchanged. This choice was made relying on evidence from previous works: first, increasing the deposition pressure and time during the growth of porous hierarchical AZO structures allows the formation of a more open forest-like material with a greater fraction of voids [19], resulting in a layer with a more irregular surface, since the morphology also evolves with increasing thickness. Second, the deposition of the compact layer is strongly influenced by the underlying porous structure [18], therefore the roughness of the top surface is directly connected with the properties of the bottom layer. The details of deposition conditions for the three sets of samples which have been studied in this work are presented in Table 1, along with the surface roughness values.

| Sample (as seen in fig.1) | (a) | (b) | (c) |
|---|---|---|---|
| Deposition pressure (Pa) | 100 / 2 | 110 / 2 | 100 / 2 |
| Thickness (nm) | 980 / 300 | 1150 / 300 | 1340 / 300 |
| Ra (nm) | 38.9 (±1.9) | 48.3 (±2.2) | 61.9 (±2.4) |
| Rms (nm) | 52.2 (±3.1) | 63.5 (±3.0) | 81.3 (±3.4) |

Table 1. List of deposition conditions for the sets of investigated samples, as reported in figure 1.

Cross-sectional SEM images of functionally graded AZO layers with increasing surface roughness are reported in figure 1.

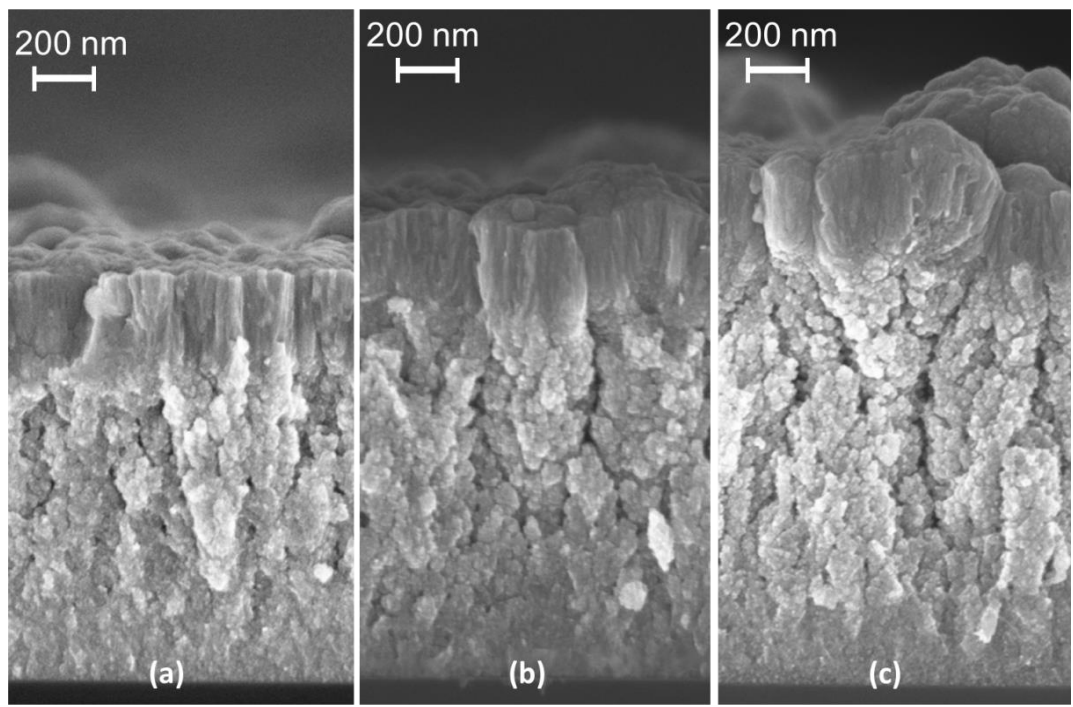

Figure 1: Cross sectional SEM images of samples grown on silicon. Sample (a) was grown with a bottom layer at 100 Pa with a 980 nm thickness, sample (b) at 110 Pa with a 1150 nm thickness, sample (c) at 110 Pa with a 1340 nm thickness.

Films are constituted by a porous layer evolving into a compact layer, and the effects of deposition time and pressure are visible both in terms of thickness (increasing both with pressure and time) and of surface morphology: in figure 1.a the compact layer is smoother and more uniformly connected, and as pressure and time are increased (1.b-c) the top layer evolves towards a more wavy and corrugated surface. This is in agreement with the AFM images presented in Figure 2, from which surface roughness values were calculated.

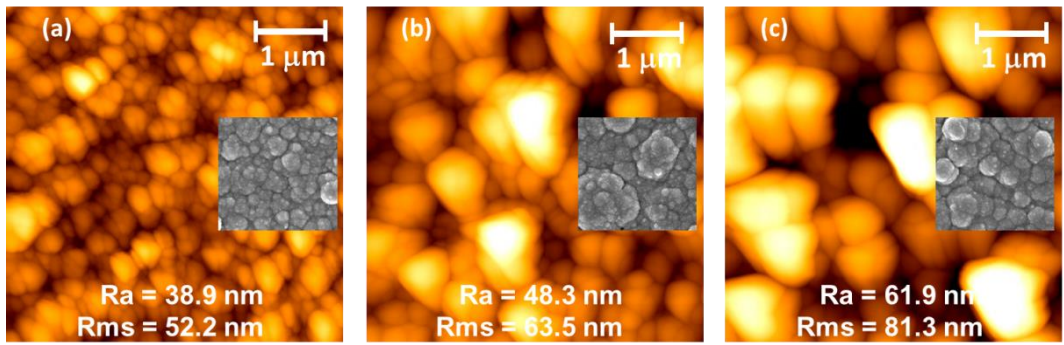

Figure 2. 5µm x 5µm tapping mode AFM images of the materials shown in figure 1. Roughness average (Ra) and root mean square (Rms) are also reported in the figure. The insets show top SEM views of the same samples at the same scale, for comparison.

Both Roughness average (Ra, mean value of the absolute displacement from average height data of on 10x10, 20x20 and 50x50 µm AFM images) and root mean square (Rms, their standard deviation) are shown.

Ra values increase from about 40 nm to about 60 nm as a result of the devised deposition conditions, and Rms values accordingly increase from 50 nm to 80 nm. Surface grain area was also calculated: results (not shown) indicate that the area of the surface domains, increasing from 0.1 µm² to 0.4 µm² with increasing surface roughness, traces the characteristic size of the underlying nanotree forest. The difference in surface grain size is clearly visible from the AFM topography images, and must not be confused with crystal domain size (previous results from XRD measurements [13-14] indicate crystal domain sizes of the order of few nanometers for single isolated porous layers and of tens of nanometers for isolated compact layers.) For comparison, SEM images of the same samples at the same scale are shown in the figure insets. We remark that roughness values on glass and silicon substrates were comparable (i.e. within the standard deviation of the data, which is reported in Table 1).

**3.2 Electrical properties**

Electrical conductivity, carrier concentration and mobility were measured in the Van der Pauw 4-point probe configuration as a function of surface roughness, in order to study the effects of surface morphology on functional properties. Results are reported in Figure 3.

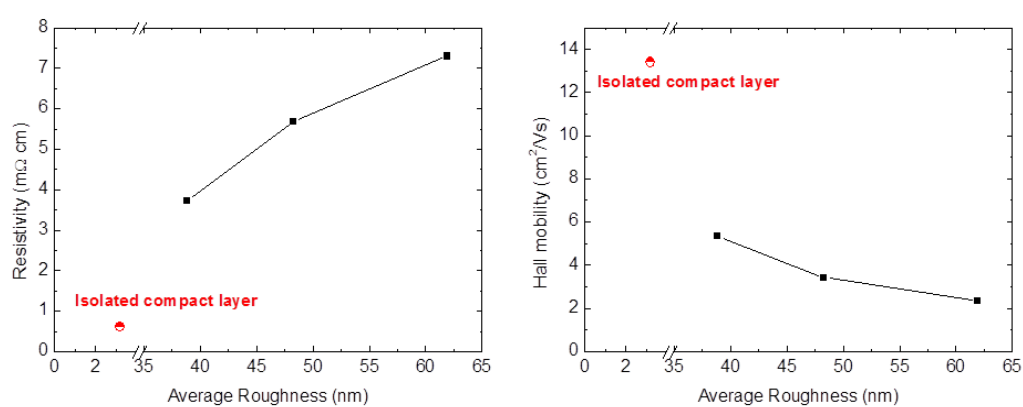

Figure 3: Electrical resistivity (left) and carrier mobility (right) as a function of average surface roughness. Resistivity is of the order of $10^{-3}$ $\Omega$ cm and increases with surface roughness, carrier mobility is of the order of few cm$^2$/Vs and decreases accordingly. Resistivity and mobility values of an isolated compact layer (red dots) are reported in correspondence of its roughness for comparison.

The values of resistivity here presented were calculated from sheet resistance using an equivalent thickness of 300 nm for the compact layer, under the hypothesis (justified by previous measurements [19]) that electrical conduction occurs in the compact layer only: this equivalent thickness corresponds to the thickness of a compact AZO layer grown in the same conditions for an equivalent amount of time, and the fairness of this choice is confirmed by SEM images (see figure 1). As for carrier (electron) mobility, the values were measured approximating the compact film to a uniform coating in order to ensure validity of the Van der Pauw hypotheses, and are hence to be considered as effective Hall mobility values.

Electrical resistivity increases with surface roughness, from $3.7 \times 10^{-3}$ Ω cm to $7.3 \times 10^{-3}$ Ω cm. Carrier mobility decreases accordingly, from 5.4 cm$^2$/Vs to 2.3 cm$^2$/Vs. The concentration of charge carriers (not shown) exhibits only small variations (from 3.15 to $3.6 \times 10^{20}$ cm$^{-3}$), pointing out a significant correlation between morphology and electrical properties, affecting especially mobility.

**3.3 Optical properties**

Optical properties were characterized in terms of transmittance and haze, i.e. scattered to total transmitted intensity ratio. The results, averaged in the visible (400 nm – 700 nm) range are presented as a function of surface roughness in figure 4.

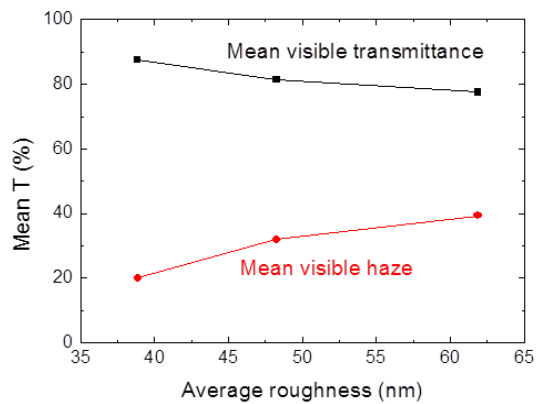

Figure 4: Optical transmittance (black squares) and haze (red dots) averaged in the 400 nm-700 nm wavelength range, as a function of average surface roughness.

Optical transmittance exhibits a small decrease with increasing surface roughness, showing mean values decreasing from almost 90% to almost 80% in the visible range. This variation can be mainly ascribed to an increase in thickness, as can be seen from the SEM images of figure 1: if the absorption coefficient (calculated as $\alpha = - \ln (T) / d$) is considered rather than the transmittance, values do not vary significantly and are of the order of 1400 cm$^{-1}$. The variations in haze are more significant: the mean value in the visible range is doubled from 20% to 40% with the variations of morphology described above in terms of porosity, thickness and surface roughness.

## 4. Discussion

Achieving control of morphology allows to investigate its effects on electrical and optical properties. The average roughness Ra value has been used as an indicator to perform a quantitative analysis, and its variations can be controlled through the morphology of the bottom porous layer alone, since the surface properties of the compact layer are strongly influenced by the underlying material, provided that the transition between the two layers occurs in a smooth and gradual way.

Results of electrical characterization (figure 3) allow to discuss the limitations to electrical conductivity in functionally graded AZO: if the growth conditions of the compact layer are left unchanged, any variation in electrical resistivity is due to morphology effects which mainly control carrier mobility. Carrier concentration in doped ZnO is influenced by intentional and unintentional dopants (besides substitutional Al which are intentional extrinsic dopants, significant contributions can be due to O vacancies or adsorption at grain boundaries, Zn interstitials, possibly H incorporation [8], [21]) and therefore variations in morphology should leave it virtually unaffected. The in-plane effective carrier mobility, on the other hand, is affected by grain boundaries (possibly smaller crystal grain size) and lack of uniform connectivity, which can increase significantly as the compact layer changes from a uniform, flat film to a wavy corrugated surface (see figure 1). The trend in mobility as a function of surface roughness allows to state that carrier mobility in graded AZO is morphology-limited, and morphology can almost fully account for the measured increase in electrical resistivity.

Optical properties also showed a significant dependence on morphology, especially for light scattering (haze). The treelike structures of the bottom layer are designed to maximize light scattering due to the proximity between their characteristic dimensions and the wavelength of visible light. While we previously demonstrated that flat compact AZO films do not give significant contribution to haze and that a more open morphology is beneficial for light scattering [19], the results of this work point to a significant dependence of haze even on small variations in morphological properties. In fact, the mean value of haze in the visible range is doubled (from 20% to 40%) with evolution of morphology and increase in thickness, and part of this increase may be also due to light deflection at the film/air interface as a consequence of surface roughness. We remark that these haze values are suitable for application in organic photovoltaic devices, as demonstrated in our previous work [18].

The variations in total transmittance, upon accounting for thickness, are negligible and allow to state that morphology is significantly influencing light scattering but not light absorption.

5. **Conclusion**

Functionally graded AZO layers were grown in different conditions to investigate the correlation between morphology and electrical/optical properties. The obtained surface roughness values ranged from 40 nm to 60 nm (Ra, or from 50 nm to 80 nm in terms of Rms) and were used as an indicator to analyze the effects of morphology on functional properties.

An increase in surface roughness was found to increase electrical resistivity due to a decrease in carrier mobility, leaving their concentration substantially unaffected and thus allowing to discuss the limitations in electrical transport due to morphology effects alone. Optical properties were also influenced by surface roughness, especially in terms of light scattering: the haze factor was doubled in correspondence with a 50% increase in surface roughness, whereas the total transmittance was not significantly altered.

Understanding these relations between morphology and functional properties can play an important role in optimization of flexible scattering transparent electrodes.